 \documentclass[sigconf]{acmart}

\usepackage{enumitem}
\usepackage{xcolor}

\usepackage{acronym}
\usepackage{array}
\usepackage{mdwmath}
\usepackage{mdwtab}

\usepackage{amssymb}
\usepackage{eqparbox}
\usepackage{amsmath}
\usepackage{tablefootnote}

\usepackage{algorithm}
\usepackage{algorithmicx}
\usepackage{algpseudocode}
\usepackage{amsmath}
\usepackage{diagbox}

\floatname{algorithm}{Algorithm}  
\newcommand{\bestperf}[1]{\textbf{\color{blue}#1}}
\newcommand{\secperf}[1]{\underline{\color{black}#1}}

\AtBeginDocument{%
  \providecommand\BibTeX{{%
    \normalfont B\kern-0.5em{\scshape i\kern-0.25em b}\kern-0.8em\TeX}}}

\setcopyright{acmcopyright}
\copyrightyear{2023}
\acmYear{2023}
\acmDOI{XXXXXXX.XXXXXXX}

\acmConference[WWW'23]{the 2023 ACM Web Conference}{April 30 – May 4, 2023}{Austin, Texas, USA}

\acmPrice{15.00}
\acmISBN{978-1-4503-XXXX-X/18/06}

\begin{document}

\title{Encoding Node Diffusion Competence and Role Significance \\ for Network Dismantling}


\author{Jiazheng Zhang}
\email{jiazhengzhang@hust.edu.cn}
\affiliation{%
  \institution{Huazhong University of Science and Technology (HUST)}
  \city{Wuhan}
  \country{China}
}

\author{Bang Wang}
\email{wangbang@hust.edu.cn}
\affiliation{%
  \institution{Huazhong University of Science and Technology (HUST)}
  \city{Wuhan}
  \country{China}
}
\renewcommand{\shortauthors}{Jiazheng Zhang and Bang Wang}

\begin{abstract}
Percolation theory shows that removing a small fraction of critical nodes can lead to the disintegration of a large network into many disconnected tiny subnetworks. The \textit{network dismantling} task focuses on how to efficiently select the least such critical nodes. Most existing approaches focus on measuring nodes' importance from either functional or topological viewpoint. Different from theirs, we argue that nodes’ importance can be measured from both of the two complementary aspects: The functional importance can be based on the nodes' competence in relaying network information; While the topological importance can be measured from nodes' regional structural patterns. In this paper, we propose an unsupervised learning framework for network dismantling, called DCRS, which encodes and fuses both node \underline{d}iffusion \underline{c}ompetence and \underline{r}ole \underline{s}ignificance. Specifically, we propose a graph diffusion neural network which emulates information diffusion for competence encoding; We divide nodes with similar egonet structural patterns into a few roles, and construct a role graph on which to encode node role significance. The DCRS converts and fuses the two encodings to output a final ranking score for selecting critical nodes. Experiments on both real-world networks and synthetic networks demonstrate that our scheme significantly outperforms the state-of-the-art competitors for its mostly requiring much fewer nodes to dismantle a network.
\end{abstract}



\maketitle

\section{Introduction}
\label{sec:Intro}
Diverse real-world systems can be abstracted as complex networks, each of which can be represented by a graph $\mathcal{G} = (\mathcal{V},\mathcal{E})$ with the node set $\mathcal{V}$ and the edge set $\mathcal{E}$. For example, a node $v_i \in \mathcal{V}$ in a computer network can be a router; while an edge $e_{ij}$ symbolizes a fiber link connecting two routers $v_i$ and $v_j$. As functional components, failures of nodes and/or edges could severely deteriorate the system normal operations. A typical example was the Italian nation-wide blackout on September 28 2003 due to cascading failures of power stations starting from a single powerline~\cite{Crucitti:et.al:2004:PhysicalA}. Another recent example was the outage of the Internet service of the American operator Century Link~\cite{Goodwin:Jazmin:2020:CNN} on August 30, 2020 arising from a single router malfunction.

\par
In the domain of network science~\cite{Boccaletti:Latora:2006:Phy.Rep., Strogatz:2001:Nature, Albert:et.al:2000:Nature}, \textit{percolation theory}~\cite{Almeira:et.al:2020:PhyRevE, Radicchi:2015:NauturePhysics} has shown that the failure of a fraction of nodes would not only cause the damage of their local structures, but also introduce some \textit{cascading failure}, even leading to the collapse of a whole network. An interesting observation is a kind of \textit{phase transition} phenomenon, that is, the network connectivity would experience an abrupt change when the number of simultaneously failed nodes reaching some threshold. These findings not only have been reported by physics scientists, but also have motivated researchers in other domains about the following network dismantling problem, that is, \textit{how to particularly attack a set of nodes to dismantle a network into many disconnected tiny subnetworks}.

\par
Following the literature~\cite{Liu:Wang:2022:NeuralNetworks, Wandelt:et.al:2018:ScientificReports, Braunstein:et.al:2016:PNAS}, the \textit{network dismantling} (ND) problem can be formally defined as to find a \textit{target attack node set} (TAS), denoted by $\mathcal{V}_t$ ($\mathcal{V}_t \subseteq \mathcal{V}$), for its removal leading to the disintegration of a network into many disconnected components, among which the \textit{giant connected component} (GCC), denoted as $\mathcal{G}_t$, is smaller than a predefined threshold $\Theta$. Formally, the objective of network dismantling is to find such a TAS $\mathcal{V}^* \subseteq \mathcal{V}$ with the minimum size, that is,
\begin{equation}\label{Eq:NDObjective}
	\mathcal{V}^* \equiv \min \big \{\mathcal{V}_{t} \subseteq \mathcal{V} : | \mathcal{G}_t |/ |\mathcal{G}| \leq \Theta \big \}.
\end{equation}
The ND problem has been proven as non-deterministic polynomial hard (NP-hard)~\cite{Braunstein:et.al:2016:PNAS}.

\par
For its NP-hardness, many heuristic solutions have been proposed for dismantling large networks~\cite{Mugisha:Zhou:2016:PhyRevE, Fan:et.al:2020:NatureML, Grassia:et.al:2021:NatureComm.}. Most of them employ some technique to measure and rank the importance of a node to network topological stability and select the top-ranked nodes into a TAS until reaching the GCC threshold. For example, many network centrality measures have been used to rank nodes, like the degree~\cite{Albert:et.al:2000:Nature}, betweeness centrality~\cite{Freeman:Linton:1977:Sociometry}, closeness centrality~\cite{Bavelas:Alex:1950:JASA}, harmonic centrality~\cite{Paolo:et.al:2014:IM}, eigenvector centrality \cite{Bonacich:et.al:1987:AmericaJS} and etc. In the last two years, a few studies have employed deep neural networks to learn node representations for node ranking and network dismantling~\cite{Liu:Wang:2022:NeuralNetworks, Zhang:Wang:2022:CIKM, Fan:et.al:2020:NatureML}. These ND-oriented neural networks have reported to achieve better performance over traditional centrality-based approaches.

\par
Inspired by the recent successes, we would like to further investigate the potentials of neural networks for the ND problem. We note that both the ND-oriented neural models of NEES~\cite{Liu:Wang:2022:NeuralNetworks} and NIRM~\cite{Zhang:Wang:2022:CIKM} mainly focus on encoding nodes' topological characteristics yet with few attention on encoding their functional capabilities. Instead, we argue that both should be encoded for ranking nodes to more efficiently dismantle a network. In this paper, we argue that network robustness can be inspected from two aspects: \textit{functional stability} and \textit{topological integrity}. We need to encode both functional capability of a node and its topological characteristics for node ranking and network dismantling. We further elaborate our motivations as follows.

\begin{figure}[t]
	\centering
	\includegraphics[ width=0.80\linewidth, height=0.8\linewidth]{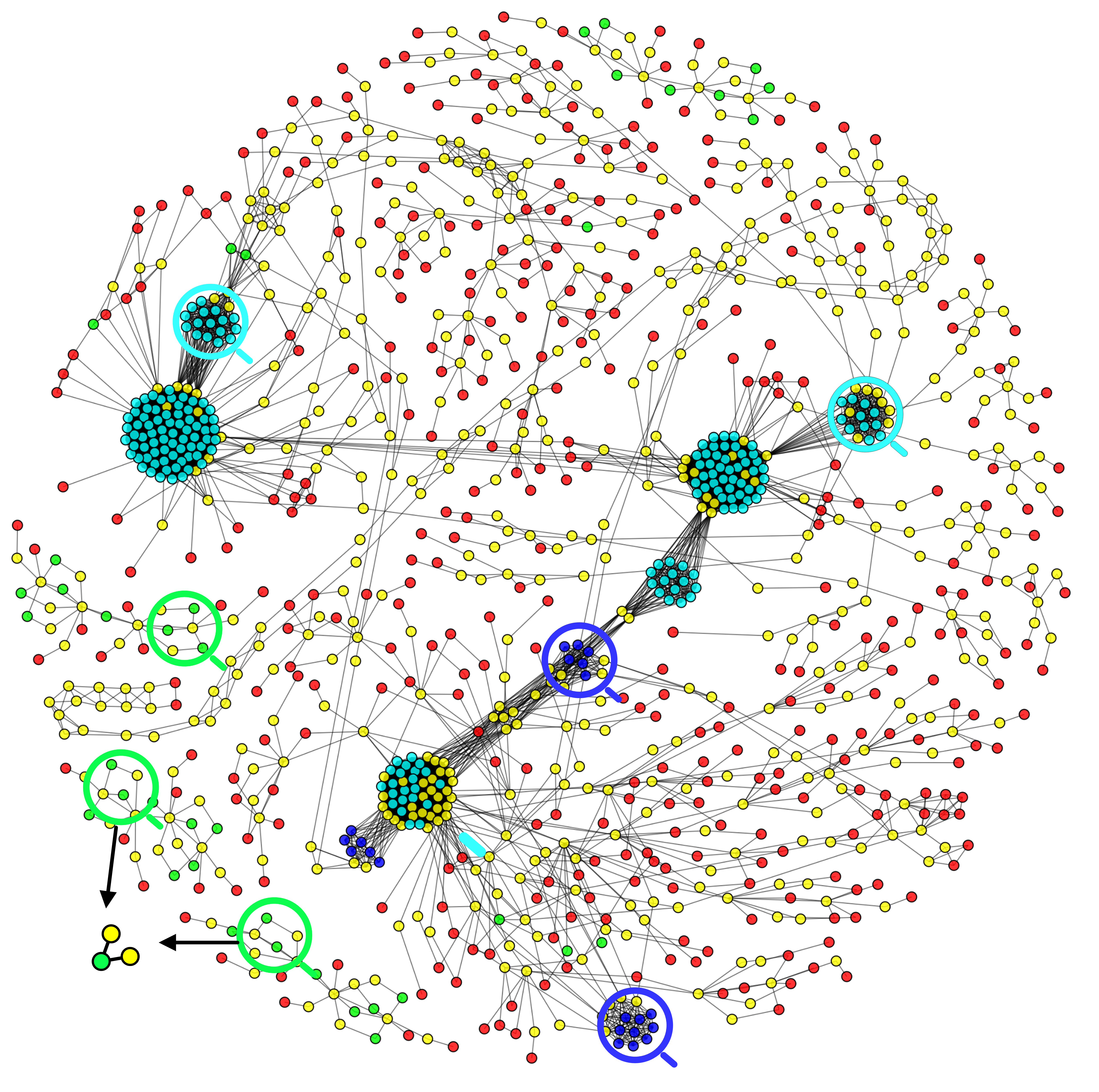}
	\caption{Illustration of role discovery by RoIX and visualized by GraphVIS~\cite{Ryan:Nesreen:2015:AAAI}. Nodes of the same role are with the same color. It can observed that many \textcolor{red}{red} nodes are kinds of peripheral nodes and a \textcolor{green}{green} node tends to connect to two \textcolor{yellow}{yellow} nodes. Some \textcolor{blue}{blue} nodes (\textcolor{cyan}{cyan} nodes) tend to form a kind of closely connected cliques that are farther apart with each other in the original network.
	}
	\label{Fig:PowerRole}
\end{figure}

\par
We regard the functional capability of a node as its \textit{diffusion competence}, e.g., its competence in relaying information flows in computer networks (or vehicle flows in road networks, etc.). It is well-known that information transmission often follows the rule of shortest-path from a source to a destination for the sake of efficiency. The network-wide functional efficiency can be well maintained, if the shortest-path between any source-destination pair is small enough. Indeed, the betweenness centrality (BC) is based on such philosophy,\footnote{The Betweenness Centrality (BC) measures the importance of a node by the ratio of number of shortest-paths passing through a node to network-wide all shortest-paths. The BC computation is often with prohibitive cost especially in large networks due to the network-wide shortest-path computation. } and a node with a high BC value suggests its pivotal importance for network-wide information diffusion. Inspired by the BC, we argue that the diffusion competence of a node can be measured from its importance for relaying information. Unlike the BC, we propose a \textit{graph diffusion network} (GDN) to encode node diffusion competence.

\par
We regard the \textit{topological characteristics} of a node as its \textit{role significance} in the whole network, which is not only determined by its \textit{egonet structural pattern} but also the significance of such an egonet pattern compared with others. We refer the \textit{egonet} of a node as its one-hop neighbors together with their edges connecting to the node. It is common that many egonets share with similar or even the same structure, such as the star and triangle structure. On the one hand, the egonet pattern often indicates the importance of the node to its regional structure stability, like the removal of a star center resulting disconnected neighbors in the egonet. Considering many possible egonet patterns, we can group them into only a few roles. On the other hand, two nodes of the same role, i.e., with similar egonet patterns, may be farther away to each other in a network, which makes it difficult to further distinguish their significance to the topological integrity of the whole network. In this paper, we propose to construct a kind of \textit{role graph} on which to encode role significance for each node. Fig.~\ref{Fig:PowerRole} illustrates the roles discovered by the RoIX algorithm~\cite{Henderson:et.al:2012:KDD}, where the aforementioned phenomena can be observed.

\par
In light of the above, we propose a network representation learning framework, called \textbf{DCRS}, to encode each node \underline{D}iffusion \underline{C}ompetence and \underline{R}ole \underline{S}ignificance for scoring and ranking node importance to network robustness, by which a TAS is constructed for network dismantling. In particular, we design a GDN neural module to encode each node a diffusion competence representation that will be converted to its \textit{diffusion score} via a multi-layer perceptron (MLP) network. We construct a role graph based on nodes' role similarities, on which we adopt a \textit{Graph Convolutional Network} (GCN) to encode each node a role significance representation and another MLP for converting to its \textit{role score}. Finally, we use a gate unit to fuse the two scores into a \textit{dismantling score} for each node. Experiments on both real-world networks and synthetic networks validate the superiority of our DCRS over the state-of-the-art competitors in terms of smaller TAS size in most cases.

\par
The remainder is organized as follows: Section~\ref{sec:Related} reviews the related work. Our DCRS scheme is presented in Section~\ref{sec:Method} and experimented in Section~\ref{sec:Experiment}. The paper is concluded in Section~\ref{sec:conclu} with some discussions.

\section{Related work}
\label{sec:Related}
\subsection{Node Centrality-based Dismantling}
Most existing solutions of estimating node importance is to first compute some centrality metric for each node, and then rank nodes and select the top-$K$ important nodes to form a TAS. Degree centrality (DC)~\cite{Albert:et.al:2000:Nature} counts the number of neighbors, and Collective Influence (CI)~\cite{Morone:et.al:2015:Nature} is proposed as an improved version of DC, considering its multi-hop neighbors' degrees. Some path-based centralities focus on global topological properties. Betweeness centrality (BC)~\cite{Freeman:Linton:1977:Sociometry} evaluates the importance of relaying information flows, and Closeness centrality (CC)~\cite{Bavelas:Alex:1950:JASA} describes the average distance of the node from other nodes. Harmonic Centrality (HC)~\cite{Paolo:et.al:2014:IM} is the sum of the reciprocal of the shortest path distances from all other nodes to the given node. In addition, Eigenvector centrality (EC)~\cite{Bonacich:et.al:1987:AmericaJS} and PageRank (PR)~\cite{Page:Brin:1999:Stanford} belong to the iterative refinement centralities: the former is determined by the number and importance of neighbors; While the latter applies the random walk with restart to obtain convergence probabilities as node importance measures.

\begin{figure*}[t]
	\centering
	\includegraphics[scale = 1.5,width=0.95\textwidth]{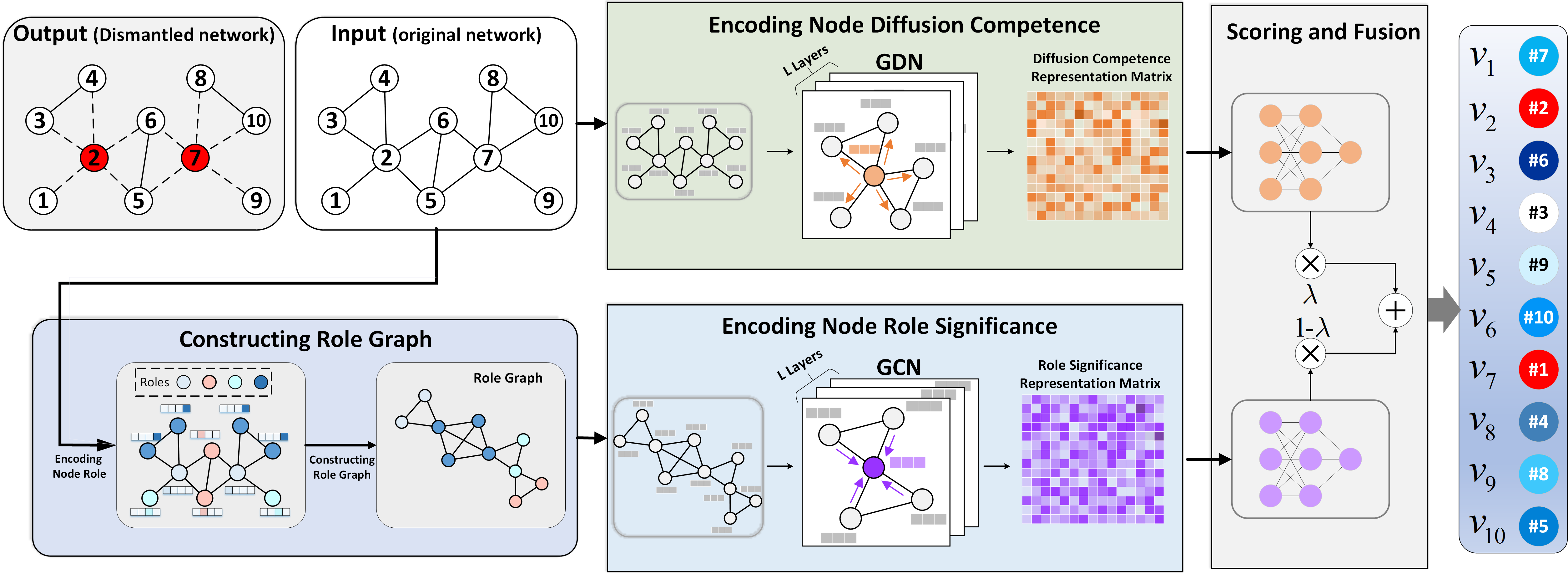}
	\caption{The DCRS framework: The left bottom part is the process of role grph construction, the right part illustrates the whole framework, including (1) encoding node diffusion competence, (2) encoding node role significance, (3) scoring and fusion.}
	\label{Fig:DCRS}
\end{figure*}

\subsection{Node Encoding-based Dismantling}
Recently, many deep neural networks have been proposed to learn node embedding for various downstream tasks, like the node classification and link prediction~\cite{Henderson:et.al:2012:KDD, Ribeiro:Saverese:2017:KDD, Rossi:Ahmed:2018:WWW, Zhang:Kou:2022:ICML}. Although they are not designed for the ND task, we can convert nodes' embeddings into dismantling scores for ranking and selecting nodes by some appropriate objective function. We will experiment a few classic network embedding schemes with our objective function.

\par
In the last two years, a few ND-oriented neural models have been designed, including the FINDER~\cite{Fan:et.al:2020:NatureML}, NEES~\cite{Liu:Wang:2022:NeuralNetworks} and NIRM~\cite{Zhang:Wang:2022:CIKM}. The FINDER adopts a reinforcement learning framework and enables a kind of \textit{reinsertion} operation, that is, heuristically reinserting already selected attack nodes into the original network. This paper focuses on the one-pass network dismantling task without reinsertion, that is, all nodes in a TAS are considered to be attacked at the same time. Both the NEES and NIRM are designed for one-pass network dismantling: The NEES~\cite{Liu:Wang:2022:NeuralNetworks} hierarchically merges some compact substructures to convert a network into a coarser one with fewer nodes and edges, during which nodes’ importance to network robustness are also learned. The NIRM~\cite{Zhang:Wang:2022:CIKM} designs a neural network to encode nodes' topological characteristics, which is trained by tiny synthetic network instances with their novel label propagation algorithm. We will compare their performance with ours in the experiment section.

\section{Methodology}
\label{sec:Method}

\subsection{Overview}
We design the DCRS framework to encode node diffusion competence as its functional importance and node role significance as topological importance to network robustness. The DCRS takes the adjacency matrix $\mathbf{A}\in \mathbb{R}^{N\times N}$ of a network as input, and outputs a vector of \textit{dismantling scores} $\mathbf{s}^{dis} \in \mathbb{R}^{N\times 1}$ for all nodes. Fig.~\ref{Fig:DCRS} illustrates the DCRS framework, which consists of the following modules: (1) encoding node diffusion competence, (2) encoding node role significance, (3) scoring and fusion.

\begin{figure}[t]
	\centering
	\includegraphics[ width=0.95\linewidth]{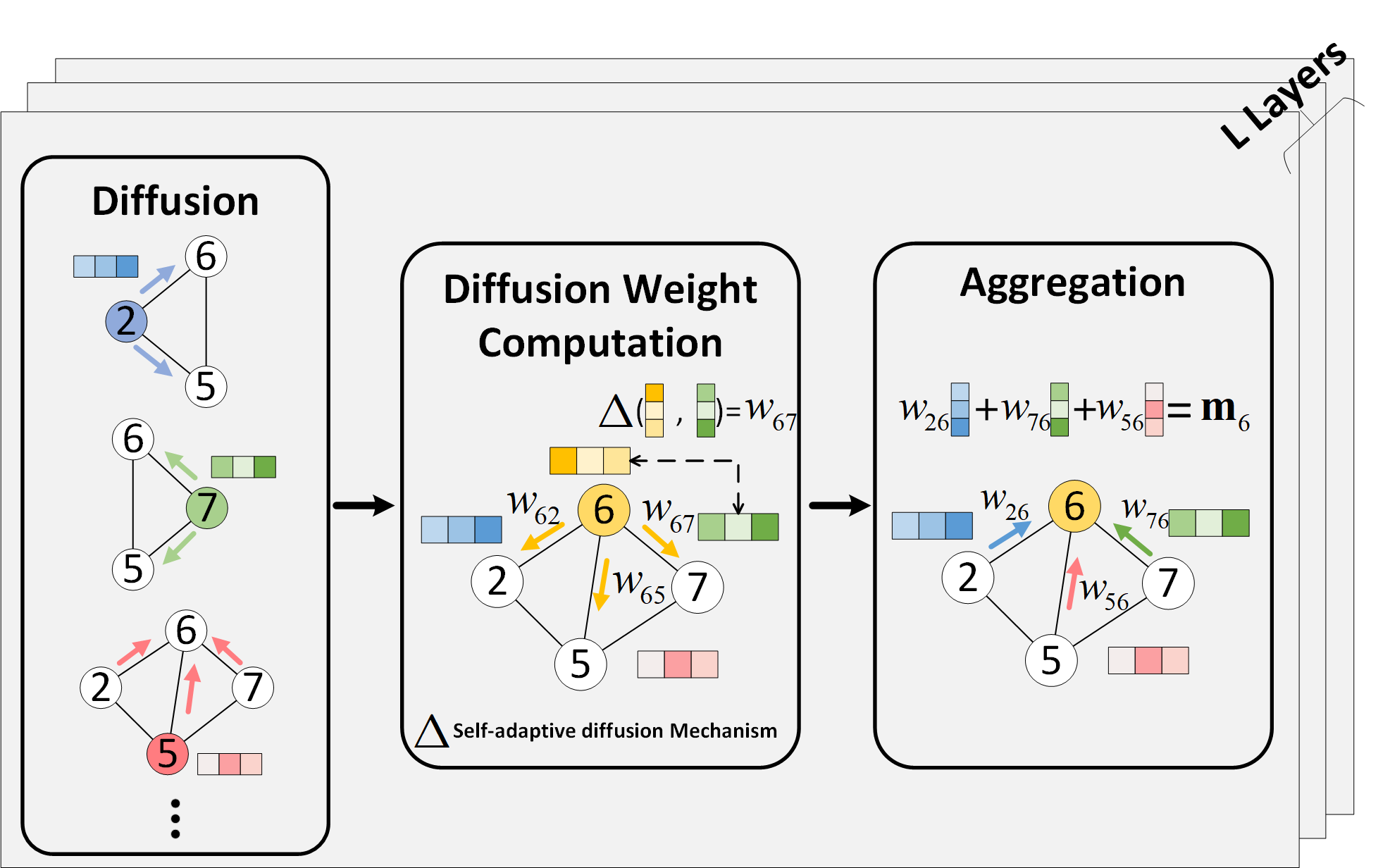}
	\caption{Illustration of graph diffusion network. Take node $v_2$, $v_5$, $v_6$, $v_7$ as example, the directed diffusion weight is assigned to the out-edge. Specifically, $v_6$ assigns the diffusion weight $w_{62}$, $w_{65}$ and $w_{67}$ (yellow edges). In the neighbor aggregation operation, $v_6$ receives its neighbor signals, which are computed as the sum of the product of the in-edge weights and the corresponding embeddings.
	}
	\label{Fig:GDN}
\end{figure}

\subsection{Encoding Node Diffusion Competence}
Node diffusion competence is used to evaluate the functional capability of a node in relaying network flows. We propose a Graph Diffusion Network (GDN) consisting of $L$ \textit{graph diffusion layers} to encode node diffusion representation.

\par
Fig.~\ref{Fig:GDN} illustrates a graph diffusion layer, which is implemented by a self-adaptive diffusion mechanism and neighbor aggregation operation. Each node $v_i$ is first initialized with an all-one vector as its diffusion representation $\mathbf{h}_i$ that can be treated as a \textit{message} to be diffused to its neighbors. For node $v_i$, we use in-edge $e_{ij}$ to describe the edge from $v_i$ to $v_j$, and out-edge the edge from $v_j$ to $v_i$. We note that this is only for ease of presentation to split one unidirectional edge into two directional edges.

\par
The first step is to diffuse the message of a node to its neighbors along the node out-edges. We propose a self-adaptive mechanism to compute the \textit{diffusion weight} $w_{ij}$ for an out-edge $e_{ij}$ of node $v_i$ as follows:
\begin{eqnarray}
	\alpha_{ij} & = & \boldsymbol\beta^{\mathsf{T}}\text{LeakyReLU}\left( \mathbf{W}_1 \mathbf{h}_i + \mathbf{W}_2 \mathbf{h}_j \right),\\
	w_{ij} & = & \text{softmax}_{j \in \mathcal{N}(i) } ( \alpha_{ij} ),
\end{eqnarray}
where $\mathcal{N}(i)$ is the set of one-hop neighbors of node $v_i$, and $\mathbf{W}_1$, $\mathbf{W}_2$, and $\boldsymbol\beta$ are learnable parameters.

\par
The second step is for each node $v_i$ to aggregate its neighbors' messages $\mathbf{h}_j \in \mathcal{N}(i)$ along its in-edge $e_{ji}$ as follows:
\begin{equation}
	\mathbf{h}_i = \sigma \left( \mathbf{W}_3 \mathbf{m}_i^{\mathsf{T}} + \mathbf{b}_3 \right),
	\quad \mathrm{where} \; \mathbf{m}_i = \sum_{v_j \in \mathcal{N}(i)} w_{j i} \mathbf{h}_j
\end{equation}
where $\mathbf{W}_3$ and $\mathbf{b}_3$ are learnable parameters to transform aggregated embedding $\mathbf{m}_i$, and $\sigma \left( \cdot \right)$ represents the activation function. We note that the diffusion weight $w_{j i}$ of the in-edge $e_{j i}$ of a node $v_i$ is computed via its neighbor $v_j$.

\par
We can stack $L$ graph diffusion layers to encode high-order neighbors' diffusion representations, which can be likened as the cascaded propagation of message transmissions in a network. After diffusion encoding, we use $\mathbf{H}$ to denote the node diffusion competence representation matrix.

\subsection{Encoding Node Role Significance}
Node role significance is used to evaluate the topological characteristics to network robustness. Starting from the egonet of a node, we apply the RoIX~\cite{Henderson:et.al:2012:KDD} to classify nodes into a few roles. Based on the node role, we propose to construct a \textit{role graph} on which to encode role representation for each node.

\subsubsection{Constructing Role Graph}
For each node and its egonet, we first design a few of local structural features, including its degree, cluster coefficient, sum of egonet nodes' degrees, ratio of edges within and leaving the ego-net, and apply the RoIX algorithm~\cite{Henderson:et.al:2012:KDD} to classify nodes into a few roles, each of which indicates nodes with similar regional structural characteristics.

\par
We construct a role graph for the input network based on the nodes' role embeddings $\mathbf{R}$ from the RoIX algorithm, which recursively aggregates nodes' local structural features into a feature matrix $\mathbf{F}$ and automatically find a rank $r$ approximation $\mathbf{R} \mathbf{M} \approx \mathbf{F}$ where $r$ is the number of roles, which is implemented by the following \textit{non-negative matrix factorization} (NMF):
\begin{equation}
	\min _{\mathbf{R}, \mathbf{M}} \lVert \mathbf{F}-\mathbf{R M}\rVert_{F}^{2}, \quad \text {s.t.}\; \mathbf{R}, \mathbf{M} \geq 0,
\end{equation}
where $\lVert \cdot \rVert_{F}$ is the Frobenius norm, the role contribution matrix $\mathbf{M} \in \mathbb{R}^{r \times d}$ represents the correlation of each role to regional structural features. The role embedding matrix $\mathbf{R} \in \mathbb{R}^{N \times r}$ quantitatively describes the mixed-membership probability of each node to the $r$ discovered roles, and $r$ is determined by the \textit{minimum description length criterion}.

\par
Fig.~\ref{Fig:DCRS} illustrates an example where the nodes in a toy network can be divided into four roles according to the RoIX algorithm. A node $v_i$ can be classified into the $k$-th role for its $k$-th element the maximum in its role embedding $\mathbf{R}_{i}$. However, we note that the semantics of roles are not specified by the RoIX algorithm, as well as the role importance to regional structure stability not evaluated. Furthermore, it is not uncommon that nodes are with the same role are not close to each other in a network, where their importance to the whole network stability are also not clear. For such considerations, we propose to construct the following role graph on which to encode node role significance to network topological stability.

\par
We construct the role graph $\mathcal{G}_r$ based on the role embedding matrix $\mathbf{R} \in \mathbb{R}^{N \times N}$. It is worth of highlighting that this operation enables nodes with similar regional structure characteristics to be connected in the role graph. In particular, we compute the role similarity between the role embeddings of node pairs. The role similarity $s_{i j}$ for the node $v_i$ and $v_j$ is computed by
\begin{equation}
	s_{i j}=\frac{\langle\mathbf{R}_{i}, \mathbf{R}_{j} \rangle}{ \lVert\mathbf{R}_{i}\rVert \lVert \mathbf{R}_{j}\rVert},
\end{equation}
where $\langle \cdot, \cdot \rangle$ stands for the dot product of two vectors, $\mathbf{R}_{i}$ ($\mathbf{R}_{j}$) the role embedding of $v_i$ ($v_j$). For each node $v_i$, we choose its top $k$ role-similar nodes, each of which is assigned an edge linking to $v_i$. After constructing the role graph $\mathcal{G}_r$, we use $\mathbf{A}_{r}$ and $\mathbf{D}_r$ to denote its adjacency matrix and diagonal degree matrix, respectively.

\subsubsection{Learning Role significance}
Fig.~\ref{Fig:DCRS} presents the constructed role graph for the toy network. Note that not only the nodes of the same role are connected, but also those nodes with similar role encodings are connected in the role graph, even some of them not connected in the original network. As role encoding is based on egonet structural features, those nodes with similar regional structures in the original network can be connected in the role graph, on which their role significance can be encoded and distinguished.

\par
In order to further distinguish roles, we propose to use a \textit{Graph Convolutional Network} (GCN) to encode the \textit{role significance} for each node in the role graph. As nodes in the role graph $\mathcal{G}_r$ are not directly related to their structural information in the original network $\mathcal{G}$, we input the GCN with all-one vectors and encode the role significance of a node as its capability of relaying and aggregating messages in the role graph.

\par
Let $\mathbf{Z}$ denote the role significance matrix. The operation of one layer GCN is as follows:
\begin{equation}
	\mathbf{Z}=\sigma ( \tilde{\mathbf{D}}_{r}^{-\frac{1}{2}} \tilde{\mathbf{A}}_{r} \tilde{\mathbf{D}}_{r}^{-\frac{1}{2}} \mathbf{Z} \mathbf{W}_{r} ),
\end{equation}
where $\sigma \left( \cdot \right)$ denotes the activation function, $\mathbf{W}_{r}$ the learnable model parameter matrix, $\tilde{\mathbf{D}}_{r} = \mathbf{D}_{r} + \mathbf{I}$ and $\tilde{\mathbf{A}}_{r} = \mathbf{A}_{r} + \mathbf{I}$ with $\mathbf{I} \in \mathbb{R}^{N \times N}$ the identity matrix. We can also stack $L$-layer of GCN to obtain the final role significance representation matrix $\mathbf{Z}$.

\subsection{Scoring and Fusion}
So far, we have obtained the node diffusion competence representation $\mathbf{H}$ and role significance representation $\mathbf{Z}$, which are designed to encode the node importance to the functional stability and structural stability of the original network. We next evaluate the \textit{diffusion competence score} $s_i^{dc}$ and the \textit{role significance score} $s_i^{rs}$ for each node $v_i$ by two multi-layer perceptron (MLP) units respectively:
\begin{gather}
s_i^{dc} = \operatorname{ReLU} \left( \mathbf{W}_4 \mathbf{H}^{\mathsf{T}}_{i} + \mathbf{b}_4 \right), \\
s_i^{rs} = \operatorname{ReLU} \left( \mathbf{W}_5 \mathbf{Z}^{\mathsf{T}}_{i} + \mathbf{b}_5 \right),
\end{gather}
where $\mathbf{W}_{4}$, $\mathbf{b}_{4}$, $\mathbf{W}_{5}$ and $\mathbf{b}_{5}$ are the learnable model parameters. As the dismantling is for the whole network, it is necessary to combine both the scores to evaluate the importance of a node for their global ranking. We design a fusion gate to get the \textit{dismantling score} $s_i^{dis} \in \mathbb{R}^{N}$ for each node by
\begin{equation}
s^{dis}_{i} = \sigma(\lambda \times s^{dc}_{i} +(1 - \lambda) \times s^{rs}_{i}),
\end{equation}
where $\lambda \in [0,1]$ is a hyper-parameter to balance the two scores and $\sigma$ denotes the sigmoid function to normalize the scores $s^{dis}_{i} \in [0, 1]$. The dismantling score $s_i^{dis}$ can be interpreted as the probability that the node $v_i$ belongs to the target attack set $\mathcal{V}_t$.

\par
Finally, We select the top-$K$ nodes with the highest dismantling scores to form the target attack set $\mathcal{V}_t$.

\subsection{Objective Function}
The objective the network dismantling is to remove the least number of nodes while causing the largest number of disconnected subnetworks each with a small enough size. Considering this, we define the loss function as follows:
\begin{equation}\label{Eq:ExpectedLoss}
	\mathcal{L} \triangleq \mathbb{E}[\mid \text {Uninfluenced nodes} \mid] + \gamma \mathbb{E}[\mid \text {target attack set} \mid].
\end{equation}

\par
The cascading effect indicates that the removal of a node not only can influence the connectivity of its local structure, but also can cause further cascaded failures of and through its neighbors. The first term of Eq.~\eqref{Eq:ExpectedLoss} penalizes the expected number of uninfluenced nodes after nodes' removal, which can be computed by
\begin{equation} \label{Eq:UnreachableNodes}
	\mathbb{E}[\mid \text {Uninfluened nodes} \mid] = \sum_{v_i \in \mathcal{V}} \prod_{v_j \in \mathcal{N}(i)} \frac{1}{1 + s_{j}^{dis}}.
\end{equation}
Note that $\frac{1}{1 + s_{j}^{dis}}$ is inversely proportional to the importance to network stability by a node $v_j$. We use $s_i^{enet} \equiv \prod_{j \in \mathcal{N}(i)} \frac{1}{1 + s_{j}^{dis}}$ to compute the collective effect of uninfluenced nodes for the node $v_i$ been removed through its egonet neighbors. That is, we measure the importance of the egonet for each node. Note that the smaller the $s_i^{enet}$, the more important of the node.

\par
As we also want to minimize the number of attack nodes, we use the sum of dismantling score $\sum s_i^{dis} $ as a regulation term to enforce the model to output reasonable solutions. Finally, the objective function of our DCRS model is given by
\begin{equation}
	\mathcal{L} = \sum_{v_i \in \mathcal{V}} \prod_{v_j \in \mathcal{N}(i)} \frac{1}{1 + s_{j}^{dis}} + \gamma \sum_{v_i \in \mathcal{V}} s_{i}^{dis},
\end{equation}
where $\gamma$ is a balancing coefficient for the two terms, which is set as $\gamma = 1$ in our experiments.

%

\begin{table}[t]
	\centering
    \small \linespread{0.9}
	\renewcommand{\arraystretch}{1.1}
	\setlength{\tabcolsep}{0.7mm}
	\caption{Statistical properties of real-world networks}
	\label{tab:Real-wrokd Dataset}
	\begin{tabular}{l c c c c c c }
		\hline
		Network & Category & Nodes & Edges & Density & $\langle k \rangle$ & Diameter \\
		\hline
		Chicago~\cite{Eash:Chon:1979:Transportation}            & Transport        & 12,979    & 20,627     & 0.0002         & 3.18         & 106 \\
		Europe~\cite{Vsubelj:et.al:2011:EurPyhJourB}            & Transport        & 1,039     & 1,305      & 0.0024         & 2.51         & 62    \\
		AirTraffic~\cite{Kunegis:et.al:2013:www}                & Airport     & 1,226     & 2,408      & 0.0032         & 3.93         & 17  \\
		Gnutella~\cite{Ripeanu:et.al:2002:IEEEComputing}      & Internet    & 8,717     & 31,525     & 0.0008         & 7.23         & 10 \\
		Route~\cite{Leskovec:et.al:2007:TKDD}         & Internet    & 6,474 & 13,895 & 0.0007 & 4.29 & 9             \\
		Blog~\cite{Kunegis:et.al:2013:www} & Political & 1,224 & 16,718 & 0.0223 & 27.32 & 8         \\
		FilmTrust~\cite{Guo:Yorke-Smith:2013:IJCAI} & Social & 874 & 1,309 & 0.0034 & 2.99 & 13                                \\
		LastFM~\cite{Rozemberczki:Sarkar:2020:CIKM}   & Social    & 7,624     & 27,806     & 0.0010         & 7.29           & 15       \\
		Flickr~\cite{Meng:et.al:2019:WSDM}        & Social       & 7,575     & 239,738   & 0.0084     & 63.30           & 4    \\
		BlogCatalog~\cite{Meng:et.al:2019:WSDM}   & Social       & 5,196     & 171,743    & 0.0127        & 66.11           & 4  \\
		HM~\cite{Kunegis:et.al:2013:www}          & Social       & 1,858     & 12,534    & 0.0073         & 13.49          & 14   \\
		RoviraVirgili~\cite{Guimera:et.al:2003:PhyRevE} & Email & 1,133 & 5,451 & 0.0085 & 9.62 & 8                                   \\
		DNCEmails~\cite{Kunegis:et.al:2013:www}   & Email       & 1,866     & 4,384      & 0.0025         & 4.70           & 8   \\
		HI-II-14~\cite{Thomas:et.al:2014:Cell}    & Protein          & 4,165     & 13,087     & 0.0016         & 6.28           & 11    \\
		Vidal~\cite{Rual:et.al:2005:Nature}       & Protein          & 3,133     & 6,726     & 0.0014         & 4.29			& 13      \\
		Figeys~\cite{Ewing:et.al:2007:MSB}        & Protein          & 2,239     & 6,432     & 0.0026         & 5.75           & 10    \\
		PPI~\cite{Dongbo:et.al:2003:NAR}          & Protein          & 2,224     & 6,609      & 0.0027         & 5.94           & 11    \\
		Genefusion~\cite{Hoglund:et.al:2006:Oncogene}          & Biology        & 291       & 279        & 0.0066         & 1.92           & 9     \\
		Bible~\cite{Kunegis:et.al:2013:www}           & Lexicon          & 1,773     & 9,131     & 0.0058          & 10.30          & 8   \\
		Wikibook~\cite{Kunegis:et.al:2013:www}   & Collab. & 553   & 648 & 0.0042 & 2.34 & 7                             \\
		Ca-GrQc~\cite{Leskovec:Krevl:2014:SNAP}    & Collab.    & 4,158     & 13,422    & 0.0016          & 6.46           & 17  \\
		UAI~\cite{Wang:et.al:2018:PAKDD}		   & Collab.    & 3,067     & 28,311    & 0.0060          & 18.46           & 7  \\
		\hline
	\end{tabular}
\end{table}


\subsection{Complexity Analysis}
The DCRS consists of three modules, and the time complexity of each module is analyzed as follows:

\par
\noindent \textbf{Encoding Diffusion Competence}. The self-adaptive diffusion mechanism takes $O(\lvert \mathcal{E} \rvert)$ and the neighbor aggregation operator requires $O(\lvert \mathcal{V} \rvert + \lvert \mathcal{E} \rvert )$. The time complexity is $O\left( L (\lvert \mathcal{V} \rvert + \lvert \mathcal{E} \rvert ) \right) $ and $L$ is number of GDN layers.

\par
\noindent \textbf{Encoding Role Significance}. There are four steps in this part. The first step of extracting egonet features takes  $O( f(\lvert \mathcal{E} \rvert + \lvert \mathcal{V} \rvert f) )$, where $f$ is the number of structural features. The second step of applying NMF for role embedding requires $O( \lvert \mathcal{V} \rvert f r)$ and $r$ is the number of roles. The third step is to compute the pairwise cosine similarity by MapReduce for role graph construction, which is achieved with empirical complexity of $O(\lvert \mathcal{V} \rvert^{1.14})$. The fourth step of encoding role significance requires $O\left( L \lvert \mathcal{E} \rvert \right) $, $L$ is number of GCN layers.

\par
\noindent \textbf{Scoring and Fusion}. The process of scoring and fusion needs $O(\lvert \mathcal{V} \rvert)$, after which performs node ranking with $O(\left| \mathcal{V} \right| log \left| \mathcal{V} \right|)$ to select the top-$K$ attack nodes.

\section{Experiment Results and Analysis}
\label{sec:Experiment}

\subsection{Experimented Networks}
We evaluate the effectiveness of our DCRS on both real-world networks and synthetic networks.

\begin{itemize}[leftmargin=*]
\item We collect twenty two real-world networks ranging from a variety of application scenarios, including internet, infrastructure, society, biology and etc., which reflect different network topological characteristics in practice. Table~\ref{tab:Real-wrokd Dataset} summarizes the statistics of these networks.

\item We use four widely used generative models to generate synthetic networks, including the Erd\"{o}s-R\'{e}nyi (ER)~\cite{Erdos:et.al:1960:evolution}, Barab\'{a}si-Albert (BA)~\cite{Barabasi:et.al:1999:Science}, Powerlaw-Cluster (PLC)~\cite{Holme:et.al:2002:PhyRevE} and  Watt-Strogatz (WS)~\cite{Watts:et.al:1998:Nature}. Note that the characteristics of synthetic networks generated by different models are usually quite different.
\end{itemize}
All experiments~\footnote{We release our code on Github at: https://github.com/JiazhengZhang/DCRS.} were implemented in Pytorch and conducted on an 8-core workstation with the following configurations: Intel Xeon E5-2620 v4 CPU with 2.10GHz, 32GB of RAM and Nvidia GeForce GTX 1080Ti GPU with 11GB memory.

\subsection{Competitors}
We compare our DCRS against 14 competent algorithms that all rank node importance for selecting the top-$K$ ones as attack nodes. These algorithms can be generally classified into two categories:

\textbf{Node centrality-based approaches}:
Many classic centralities can be used for network dismantling.
\begin{itemize}[leftmargin=*]
\item DC (Degree Centrality)~\cite{Albert:et.al:2000:Nature} remove nodes in the order of node degree.

\item BC (Betweeness Centrality)~\cite{Freeman:Linton:1977:Sociometry} calculates the frequency of all-pairs shortest-paths through a node.

\item CC (Closeness Centrality)~\cite{Bavelas:Alex:1950:JASA} is the reciprocal of the average shortest path distance between the node and other nodes.

\item EC (Eigenvector Centrality)~\cite{Bonacich:et.al:1987:AmericaJS} is based on the eigenvectors of the adjacency matrix.

\item HC (Harmonic Centrality)~\cite{Paolo:et.al:2014:IM} is a variant of the CC algorithm that can be applied to disconnected original network.

\item CI (Collective Influence)~\cite{Morone:et.al:2015:Nature} measures a node importance by considering the high-order local structure centered at the node.

\item PR (PageRank)~\cite{Page:Brin:1999:Stanford} is based on the random walk with restart.
\end{itemize}

\par
\textbf{Node encoding-based approaches}:
We explore the capabilities of neural networks for node encoding on the ND task.
\begin{itemize}[leftmargin=*]
\item DW (DeepWalk)~\cite{perozzi:et.al:2014:SIGKDD} is built upon the skip-gram model to learn nodes' embeddings.

\item NV (Node2Vec)~\cite{grover:et.al:2016:SIGKDD} extends DeepWalk with biased second order random walks to explore the high-oder neighbors.

\item RV (Role2Vec)~\cite{Ahmed:et.al:2020:TKDE} adopts attributed random walks to learn structural role-based embeddings.

\item GCN~\cite{kipf:et.al:2017:ICLR} is based on message passing and aggregation to learn nodes' embeddings.

\item GAT~\cite{Petar:et.al:2018:ICLR} extends the GCN by leveraging the masked self-attention mechanism for nodes' embedding.

\item NIRM~\cite{Zhang:Wang:2022:CIKM} is a state-of-the-art neural model particularly designed for the ND task, which encodes both local structural and high-order topological characteristics for scoring and ranking nodes.

\item NEES~\cite{Liu:Wang:2022:NeuralNetworks} is another state-of-the-art neural model for the ND task, which hierarchically merges compact substructures and extract essential structures to learn and fuse nodes' importance at different scales.

\end{itemize}
Note that except NIRM and NEES, others are for network representation learning. To suit the ND task, for the GCN and GAT we add a final linear layer that takes the nodes' embedding as input and outputs scores. For DW, NW and RV, we train a MLP for each of them to convert nodes' embeddings to scores. They all adopt the same objective function as our model.

\begin{table*}[thb]
	\centering
	\renewcommand{\arraystretch}{0.88}
	\setlength{\tabcolsep}{1.2 mm}
	\caption{Comparison of normalized TAS size on real-world networks (The dismantling threshold $\Theta = 0.01$). The best is marked in \bestperf{bold blue} while the second best is in \secperf{underline}. }
	\label{tab:result_Realworld}
	\begin{tabular}{l c c c c c c c c c c c c c c | c l}
		\toprule
		Datasets & DC & BC & CC & EC & HC & CI & PR & DW & NV & RV & GAT & GCN & NIRM & NEES & DCRS & Imprv.(\%) \\
		\hline
		Chicago & 50.8 & 55.58 & 98.93 & 96.49 & 97.81 & 60.42 & 51.56 & 65.98 & 70.74 & 75.24 & 77.73 & 76.65 & 46.58 & \secperf{42.15} & \bestperf{35.3} & +16.25 \\
		Europe & 41.29 & 48.99 & 97.69 & 97.88 & 97.11 & 82.19 & 33.69 & 80.65 & 46.2 & 92.49 & 87.78 & 89.51 & 38.11 & \secperf{28.1} & \bestperf{23.39} & +16.76\\
		AirTraffic & 32.79 & 50.98 & 86.13 & 92.33 & 83.69 & 68.03 & 28.14 & 87.68 & 73.82 & 96.33 & 76.59 & 98.04 & \secperf{25.61} & 26.43 & \bestperf{23.57} & +7.97\\
		Gnutella & 36.64 & 38.35 & 46.46 & 58.85 & 45.99 & 40.35 & 35.07 & 95.62 & 95.23 & 97.96 & 63.78 & 98.82 & \secperf{35.03} & 43.32 & \bestperf{32.45} & +7.37  \\
		Route & \secperf{4.06} & 4.59 & 56.86 & 60.98 & 55.11 & 6.43 	& 4.49 	& 83.58 & 93.03 	& 97.93 & 60.32 & 4.46 & 55.92 	& 4.79 & \bestperf{3.43} & +15.52 \\
		Blog & 45.34 & \secperf{44.12} 	& 68.46 	& 66.01 	& 67.97 & 52.04 & 44.53 	& 96.49 	& 96.81 	& 95.26 	& 89.05 & 98.94 	& 55.88 	& 46.73 	& \bestperf{37.09} & +15.93   \\
		FilmTrust & 22.77 	& 33.75 	& 70.02 	& 89.93 	& 70.02 	& 42.68 	& \secperf{22.20} 	& 89.13 	& 64.87 	& 88.67 	& 81.69 	& 98.74 	& 44.39 	& 25.97 	& \bestperf{19.11}  & +13.92   \\
		LastFM & 31.69 & 40.11 & 78.4 & 86.46 & 78.78 & 48.6 & \secperf{27.96} & 94.24 & 79.55 & 96.43 & 61.23 & 98.95 & 70.96 & 31.47 & \bestperf{26.02} & +6.94\\
		Flickr & 51.76 & \secperf{45.68} & 62.19 & 60.05 & 62.05 & 51.14 & 48.2 & 93.85 & 91.71 & 98.36 & 81.33 & 99 & 57.25 & 50.27 & \bestperf{42.57} & +6.81\\
		BlogCatalog & 90.22 & \secperf{82.79} & 98.52 & 98.71 & 98.4 & 97.83 & 84.47 & 98.36 & 98.36 & 98.63 & 97.13 & 99 & 97.84 & 84.91 & \bestperf{80.6} & +2.65\\
		HM & 40.53 & 40.64 & 80.41 & 82.08 & 78.47 & 57.91 & \secperf{33.48} & 84.98 & 91.07 & 96.56 & 87.62 & 98.98 & 34.98 & 39.77 & \bestperf{31.38} & +6.27\\
		RoviraVirgili	& 48.46 	& 54.55 	& 74.32 	& 79.88 	& 72.90 	& 59.31 	& \secperf{44.40} 	& 95.23 	& 94.88 	& 97.26 	& 86.41 	& 98.94 	& 45.01 	& 52.34 	& \bestperf{43.07}  & +3.00\\
		DNCEmails & 8.09 & 15.7 & 90.14 & 90.25 & 88.1 & 30.33 & \secperf{6.16} & 27.6 & 31.62 & 97.75 & 74.12 & 7.5 & 6.38 & 10.45 & \bestperf{5.89} & +4.38\\
		HI-II-14 & 17.53 & 25.74 & 60.05 & 65.14 & 58.66 & 32.73 & 16.59 & 92.05 & 75.29 & 98.06 & 69.48 & 98.99 & 17.07 & \secperf{16.81} & \bestperf{13.97} & +15.79\\
		Vidal & 20.84 & 23.46 & 60.77 & 69.55 & 59.85 & 31.6 & 20.14 & 92.82 & 89.59 & 95.34 & 54.23 & 98.82 & 31.38 & \secperf{17.3} & \bestperf{16.02} & +7.4\\
		Figeys & 18.89 & 18.89 & 50.2 & 56.9 & 51.14 & 25.9 & 16.03 & 36.67 & 86.02 & 94.6 & 54.27 & 30.64 & 39.53 & \secperf{10.05} & \bestperf{8.93} & +11.14\\
		PPI & 27.34 & 35.7 & 55.13 & 60.52 & 50.31 & 34.67 & \secperf{24.19} & 80.98 & 41.32 & 97.12 & 61.38 & 98.88 & 25.49 & 24.82 & \bestperf{21.67} & +10.42\\
		Genefusion & 19.24 & 21.65 & 79.04 & 98.97 & 81.1 & 82.82 & \secperf{13.4} & 72.85 & 78.69 & 97.25 & 76.63 & 98.28 & 13.75 & 19.59 & \bestperf{11.34} & +15.37\\
		Bible & 30.46 & 32.43 & 67.17 & 71.29 & 63.4 & 53.86 & 30.23 & 95.88 & 80.6 & 97.57 & 84.26 & 98.98 & \secperf{27.35} & 35.25 & \bestperf{26.11} & +4.53\\
		Wikibook & 14.65 	& 15.37 	& 89.87 	& 92.22 	& 91.50 	& 33.27 	& \bestperf{11.75} 	& 17.18 	& 55.15 	& 96.38 	& 43.22 	& 16.46 	& 67.63 	& 13.74 	& \secperf{11.93} 	& -1.51	\\
		Ca-GrQc & 25.44 & 24.63 & 62.31 & 65.42 & 60.63 & 39.56 & \secperf{19.1} & 70.8 & 69.36 & 97.45 & 73.02 & 98.8 & 24.29 & 19.17 & \bestperf{17.58} & +7.96\\
		UAI & 46.2 & 55 & 80.7 & 73.07 & 79.36 & 54.22 & \secperf{42.13} & 97.39 & 96.54 & 97.72 & 86.04 & 98.99 & 75.48 & 50.77 & \bestperf{40.04} & +4.96\\
		\bottomrule
	\end{tabular}
\end{table*}

\begin{figure*}[t]
\centering
\includegraphics[width=\textwidth, height=0.45\textwidth]{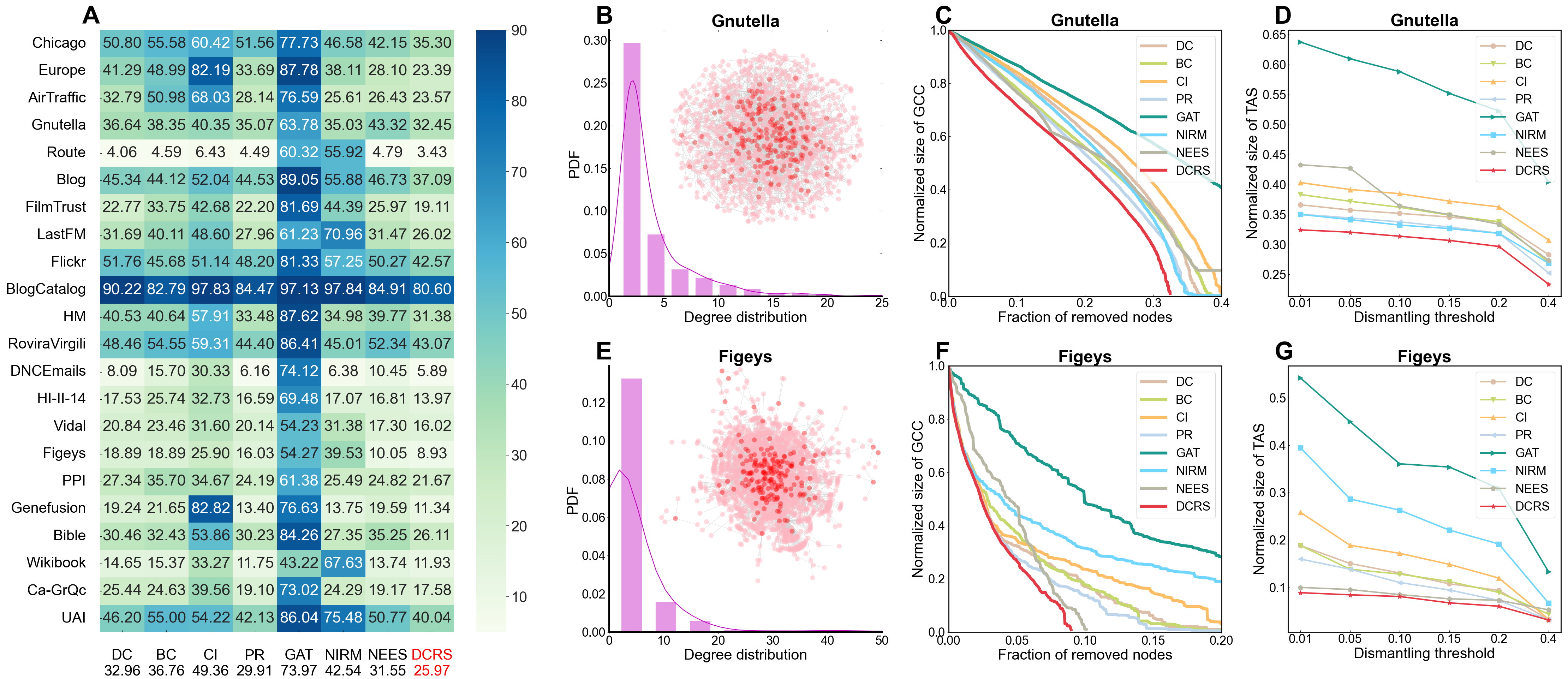}
\caption{Comparison of dismantling performance on twenty-two real-world networks.
(\textit{A}): The smaller the normalized size of TAS $\rho$, the lighter the color. Among the 22 real world networks, out DCRS achieves the best on 21 networks, and the second best on 1 network. The bottom row provides the averaged $\rho$ over all 22 networks, where ours is 25.97 and second best is 29.91. (\textit{B} and \textit{E}): the node degree distribution and the attack nodes (red) selected by our DCRS for two real world networks: Guntella containing 8,717 nodes and 31,525 edges, and Figeys containing 2,239 nodes and 6,432 edges. (\textit{C} and \textit{F}) : the normalized size of GCC when removing different fractions of target attack nodes. (\textit{D} and \textit{G}): the dismantling performance $\rho$ against different dismantling thresholds $\Theta$.}
\label{fig:Real_Experiment}
\end{figure*}

\begin{figure*}[thb]
\centering
\includegraphics[width=\textwidth, height=0.51\textwidth]{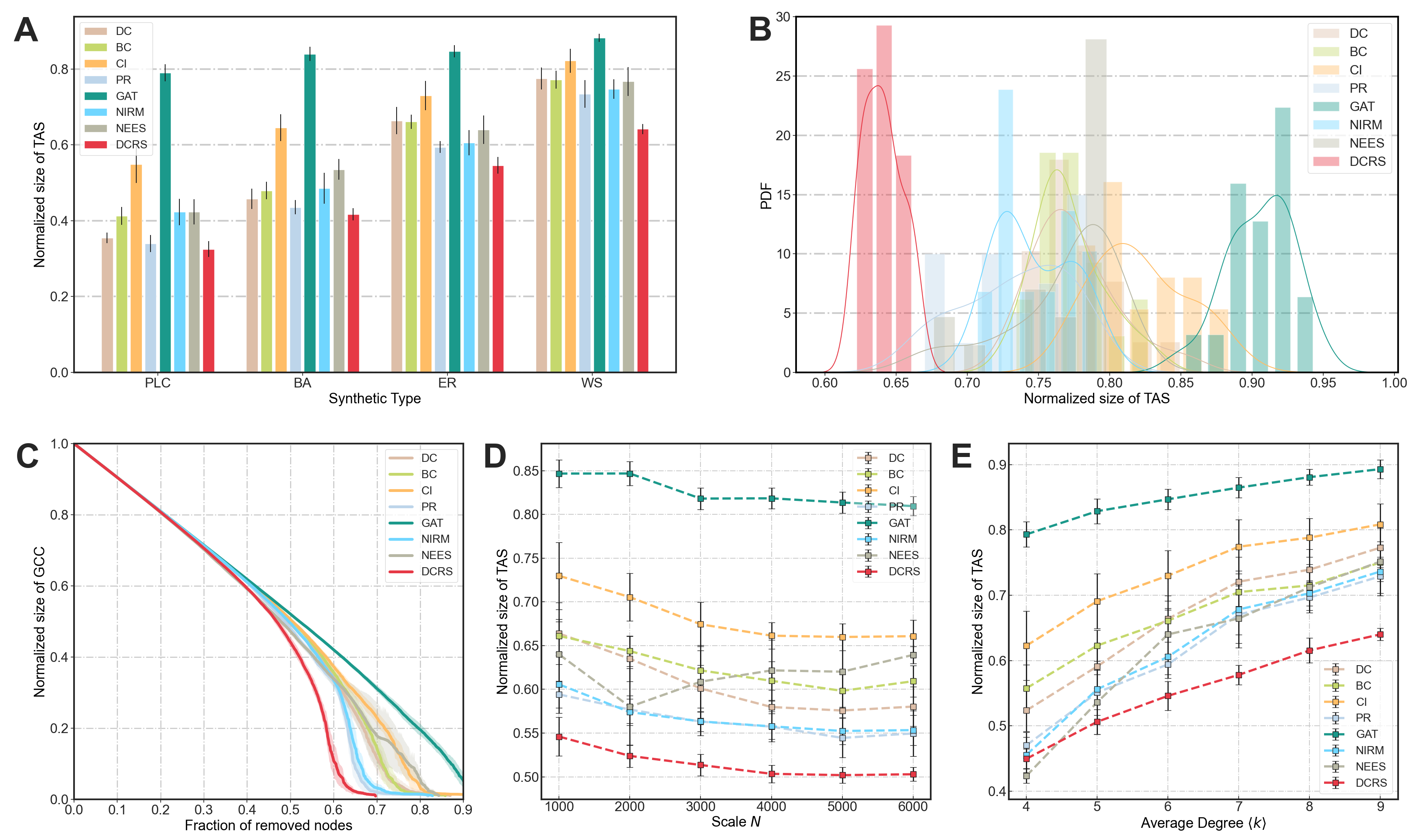}
\caption{Dismantling performance of DCRS on synthetic model networks. (\textit{A}) Comparison on four types of synthetic networks (network size $N$=1000), i.e. BA, WS, PLC and ER. Our DCRS achieves the best on four types of synthetic model networks.
(\textit{B}): the distribution of normalized size of TAS on WS ($m=8, p=0.8$) networks over 20 test instances. (\textit{C}): the normalized size of GCC when removing different fractions of target attack nodes, in which the line is the mean value and shaded area indicates the standard deviation. (\textit{D}): the dismantling performance $\rho$ on ER networks with fixed network average degree $\langle k \rangle=6$ and varying network size $N$ from 1000 to 6000. (\textit{E}): dismantling performance of normalized size of TAS on ER networks with fixed network size $ N=1000$ and varying $\langle k \rangle$.  }
\label{fig:Syn_Experiment}
\end{figure*}

\subsection{Experimental results}
\textbf{Real-world networks}: Table~\ref{tab:result_Realworld} presents the dismantling performance in terms of \textit{normalized TAS size} $\rho$ (i.e., $\rho\equiv |\mathcal{V}_t| / |\mathcal{V}|$) for twenty two real-world networks. Note that the smaller the $\rho$, the better the dismantling efficiency. From the table, we can observe that our DCRS achieves the best performance in terms of the smallest $\rho$ in 21 out of 22 real networks, showing its superiority over the competitors. In the one case of not the best, our DCRS also achieves the comparable performance with the best one. It is worth of noting that as these real-world networks are with much different statistics and characteristics, it is a rather challenging objective for one single model to excel the best in all of them. In Table~\ref{tab:result_Realworld}, the performance improvement (the rightmost column) is computed over the second-best scheme; While for different real-world networks, the second-best schemes are also different, indicating that all schemes are kinds of approximate solutions to the NP-hard network dismantling task.

\par
Fig.~\ref{fig:Real_Experiment} provides the visual comparisons for the most competitive schemes, including the DC, BC, CI, PR, GAT, NIRM, NEES and DCRS. Fig.~\ref{fig:Real_Experiment}~\textit{A} uses different colors to visualize $\rho$ for the twenty two real-world networks with sizes ranging from 291 to 12,979 nodes. The bottom line provides the averaged $\rho$ over the experimented networks, where ours is much smaller than the competitors. On average, our DCRS requires about $25.97\%$ nodes to dismantle a network; While the second best one needs removing $29.91\%$ nodes.

\par
Fig.~\ref{fig:Real_Experiment}~\textit{B} and ~\textit{E} visualize the degree distribution and target attack nodes (red) in the Gnutella and Figeys network. Figs.~\ref{fig:Real_Experiment}~\textit{C} and ~\textit{F} plot the dismantling performance in terms of the normalized GCC (NGCC) against the fraction of removed nodes in the decreasing order of their scores. It is interesting to see that our DCRS dismantles the Figeys network with only $8.93\%$ nodes, which is significantly less than others. We note that the area under the NGCC curve can also reflect the attack efficacy of a dismantling scheme. The areas of our DCRS are $1592.45$ and $70.63$ in the Gnutella and Figeys network, respectively, much smaller than the state-of-the-art methods NIRM of $1868.75$ and NEES of $98.49$ in the corresponding network. Figs.~\ref{fig:Real_Experiment}~\textit{D} and \textit{G} present that our DCRS outperforms the others in terms of the normalized TAS size under different dismantling thresholds $\Theta$ in the two networks.

\begin{table}[t]
\centering
\caption{Ablation study of dismantling performance $\rho$ when using node intermediate score and the final score.}
\footnotesize
\renewcommand{\arraystretch}{1.3}
\setlength{\tabcolsep}{0.23mm}
\begin{tabular}{l| c c c c c c c c c c}
	\hline
	Methods     & AirTraffic    & Europe & Ca-GrQc & Chicago     & UAI     & PPI     & HM          & Blog    & Film.  & Rovira.                       \\
	\hline
	\hline
	w/o RS      & 92.50         & 63.62  & 96.32   & 46.15       & 83.99   & 98.16   & 98.82       & 98.86   & 97.14  & 97.26         \\
	w/o DC      & 84.09         & 65.83  & 72.58   & 61.57       & 74.86   & 56.65   & 78.36       & 85.87   & 56.52  & 89.67      \\
	DCRS        & \textbf{23.57}      & \textbf{23.39}  & \textbf{17.58}  & \textbf{35.3}   & \textbf{40.04}  & \textbf{21.67} & \textbf{31.38}   & \textbf{37.09}  & \textbf{19.11}   & \textbf{43.07}     \\
	\hline
	Methods     & Gnutella   & HI-II-14   & Flickr & Figeys  & LastFM  & Vidal  & Bible & BlogC. & Gene. & DNC. \\
	\hline
	\hline
	w/o RS      & 91.38      & 14.19      & 99.01  & 10.09  & 97.86 & 16.63 & 99.04   & 98.77    & 11.68     & 6.48 \\
	w/o DC      & 58.07      & 65.11      & 81.93  & 58.15  & 61.31 & 50.65 & 81.16   & 85.07    & 85.57     & 63.02   \\
	DCRS         & \textbf{32.45}    & \textbf{13.97}   & \textbf{42.57} & \textbf{8.93}  & \textbf{26.02} & \textbf{16.02}  & \textbf{26.11} & \textbf{80.6}    & \textbf{11.34} & \textbf{5.89}  \\
	\hline
\end{tabular}
\label{Table:Ablation}
\end{table}

\par
\textbf{Effectiveness of scoring and fusion}: Recall that our diffusion competence score is designed for ranking node importance to network functional stability, and role significance score to network topological stability; While the two are fused for final node ranking. We can also use only one of them for network dismantling and have two variants, DCRS w/o RS and DCRS w/o DC. Table~\ref{Table:Ablation} compares the dismantling performance for the real-world networks. We can see that a single variant cannot outperform the other in all experimented networks; While the DCRS achieves the best and greatly outperforms either variant. This suggests that our design objective of encoding both functional and topological stability can be well integrated with the fusion mechanism.

\par
\textbf{Synthetic model networks}: Fig.~\ref{fig:Syn_Experiment}~\textit{A} plots the mean normalized TAS size $\bar\rho$ (and the standard deviation) for the eight most competitive schemes, where each result is averaged over 20 generated network instances. Note that a synthetic model generates network instances with similar network characteristics, but the generated instances by different synthetic models are usually with different network characteristics, like the uniform degree distribution by the ER model but the power-law degree distribution by the BA model. From Fig.~\ref{fig:Syn_Experiment}~\textit{A}, it is seen that our DCRS outperforms the competitors in all cases, which validates its applicability for the four classic synthetic models.

\par
We further take a close look at the dismantling performance on ER networks with different generation parameters. We note that although our DCRS outperforms the others in terms of smaller $\bar\rho$, it may not be the best one in every network instance.
Fig.~\ref{fig:Syn_Experiment}~\textit{B} plots the distribution of $\rho$ over experimented instances, where our DCRS achieves the smallest value in most instances and smaller variance as indicated from the fitted distribution curve. Fig.~\ref{fig:Syn_Experiment}~\textit{C} plots the NGCC against the removal fractions, where our DCRS is with the highest dismantling efficiency in terms of the smallest area under the average curve. Fig.~\ref{fig:Syn_Experiment}~\textit{D} shows that for fixing network average degree $\langle k \rangle=6$ and increasing network size $N$, our DCRS outperforms the competitors in terms of the least target attack nodes. Fig.~\ref{fig:Syn_Experiment}~\textit{E} presents similar results for fixed $N=1000$ and varying $\langle k \rangle$ from 4 to 9.

\section{Conclusion}
\label{sec:conclu}
This paper has designed a DCRS neural model for ranking and selecting attack nodes to dismantle a network. We have designed a graph diffusion neural network to encode node diffusion competence for measuring node importance to network functional stability, and have proposed to construct a role graph on which to encode node role significance for measuring node importance to network topological integrity. A new objective function has been designed for training the proposed neural model. Experiments on real-world and synthetic networks have validated the superiority of our model over the state-of-the-art competitors in terms of its mostly requiring much fewer attack nodes for network dismantling.

\par
Given its NP-hardness, we acknowledge that our solution to the network dismantling problem is still far from perfection. For example, how to encode regional structural patterns with high-order egonets and how to fast construct the role graph for a very large network still remain further investigations.


\bibliographystyle{ACM-Reference-Format}
\bibliography{DCRS}

\begin{figure*}[tb]
	\centering
	\includegraphics[width=0.85\textwidth]{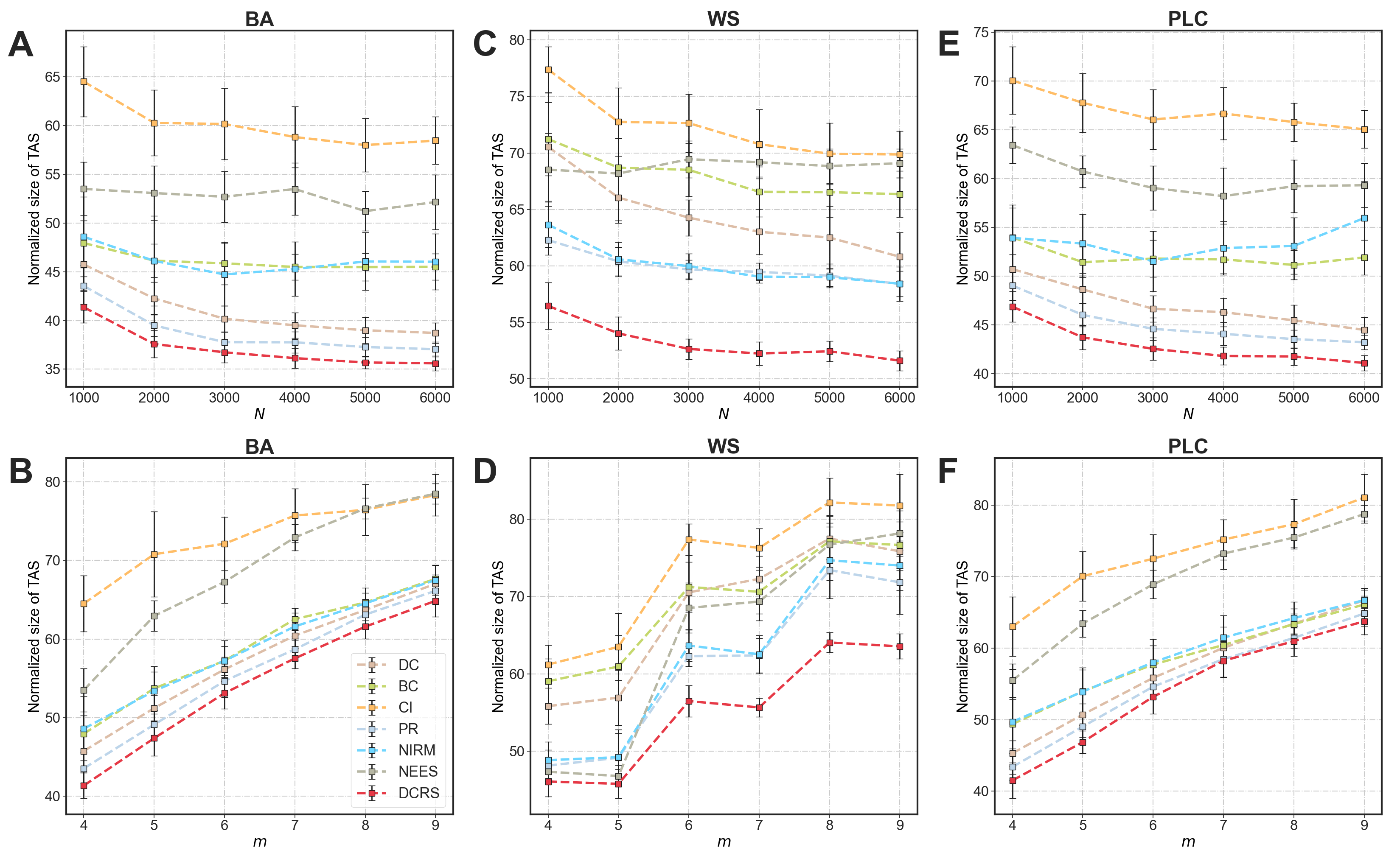}
	\caption{Performance of DCRS on different synthetic parameters. Our DCRS consistently achieves the best on different synthetic parameters. (\textit{A}) Performance on BA networks with fixed $ m=4$ and varying Scale $N$ from 1000 to 6000. (\textit{B}) Performance of normalized size of TAS on BA networks with  $ N=1000$ and varying $m$. (\textit{C}) Performance on WS networks of varying $N$ with $ m=6, p=0.8$. (\textit{D}) Performance on WS networks with $p=0.8, N=1000$ and varying $m$. (\textit{E}) Performance on PLC networks of varying $N$ with $ m=5, p=0.1$. (\textit{F}) Performance on PLC networks with $p=0.1, N=1000$ and varying $m$.}
	\label{fig:SI_Synthetic}
\end{figure*}

\begin{table*}[thb]
	\centering
	\renewcommand{\arraystretch}{0.75}
	\setlength{\tabcolsep}{1.6 mm}
	\caption{Comparison of attack efficiency on real-world networks (Area under the NGCC curve). The best is marked in \bestperf{bold blue} while the second best is in \secperf{underline}. }
	\label{tab:SI_Efficiency}
	\begin{tabular}{l c c c c c c c c}
		\toprule
		Datasets & DC & BC & CI & PR & GAT & NIRM & NEES & DCRS  \\
		\hline
		Chicago & 2764.18 & 2767.76 & 3092.84 & 2537.1 & 5030.01 & \secperf{2460.81} & 2523.84 & \bestperf{2372.75}  \\
		Europe & 74.68 & 130.36 & 107.39 & 77.1 & 358.2 & \secperf{69.66} & 91.62 & \bestperf{69.44}  \\
		AirTraffic & 163.06 & 186.6 & 235.94 & \secperf{144.69} & 471.54 & 153.42 & 165.47 & \bestperf{141.99}  \\
		Gnutella & 1970.52 & 1873.36 & 2140.82 & \secperf{1758.8} & 2927.99 & 1868.75 & 1907.3 & \bestperf{1592.45}  \\
		Route & \secperf{79.58} & 87.63 & 92.49 & 80.21 & 1189.44 & 590.11 & 114.08 & \bestperf{77.85}  \\
		Blog & 277.66 & 266.01 & 283.93 & \secperf{262.19} & 488.33 & 285.4 & 280.51 & \bestperf{242.48}  \\
		FilmTrust & 37.03 & 39.18 & 53.13 & \bestperf{32.89} & 210.93 & 41.17 & 42.99 & \secperf{34.39}  \\
		LastFM & 1247.11 & 1299.6 & 1514.48 & \bestperf{1108.67} & 2316.2 & 1444.53 & 1302.86 & \secperf{1110.11}  \\
		Flickr & 2111.5 & \secperf{1954.3} & 2120.04 & 2049.41 & 3094.37 & 2155.77 & 2153.59 & \bestperf{1898.5}  \\
		BlogCatalog & 2475.55 & \secperf{2439.34} & 2515.8 & 2457.38 & 2572.47 & 2491.8 & 2441.33 & \bestperf{2402.56}  \\
		HM & 303.61 & 311.13 & 347.81 & \secperf{283.78} & 824.57 & 296.06 & 342.91 & \bestperf{278.43}  \\
		RoviraVirgili & 284.46 & 290.6 & 305.15 & \secperf{269.79} & 512.86 & 273.73 & 305.62 & \bestperf{264.34}  \\
		DNCEmails & 27.05 & 31.89 & 33.17 & \bestperf{23.1} & 720.68 & 26.77 & 58.85 & \secperf{24.77}  \\
		HI-II-14 & 286.29 & 331.01 & 367.1 & \secperf{269.51} & 1122.37 & 283 & 319.13 & \bestperf{265.94}  \\
		Vidal & 204.95 & 231.04 & 281.15 & \secperf{200.44} & 548.01 & 217.74 & 242.06 & \bestperf{197.23}  \\
		Figeys & 102.47 & 106.01 & 128.53 & \secperf{88.93} & 346.37 & 193 & 98.49 & \bestperf{70.63}  \\
		PPI & 268.3 & 283.06 & 319.05 & \secperf{251.28} & 646.44 & 263.46 & 288.63 & \bestperf{241.4}  \\
		Genefusion & 2.21 & 3.86 & 11.27 & 2.33 & 37.75 & \bestperf{2.06} & 4.26 & \bestperf{2.06}  \\
		Bible & 232.8 & 251.46 & 274.77 & \secperf{226.96} & 685.91 & 227.97 & 304.59 & \bestperf{224.53}  \\
		Wikibook & 7.07 & \bestperf{6.41} & 10.57 & 7.23 & 138.56 & 12.51 & 8.1 & \secperf{6.71} \\
		Ca-GrQc & 538.69 & \bestperf{349.98} & 561.97 & 353.72 & 1292.11 & 533.41 & 430.67 & \secperf{353.6} \\
		UAI & 719.36 & 740.24 & 762.67 & \secperf{691.85} & 1250.52 & 768.84 & 854.37 & \bestperf{664.61} \\
		\bottomrule
	\end{tabular}
	
\end{table*}


\begin{figure*}[htb]
	\centering
	\includegraphics[width=0.85\textwidth, height= 0.36\textwidth]{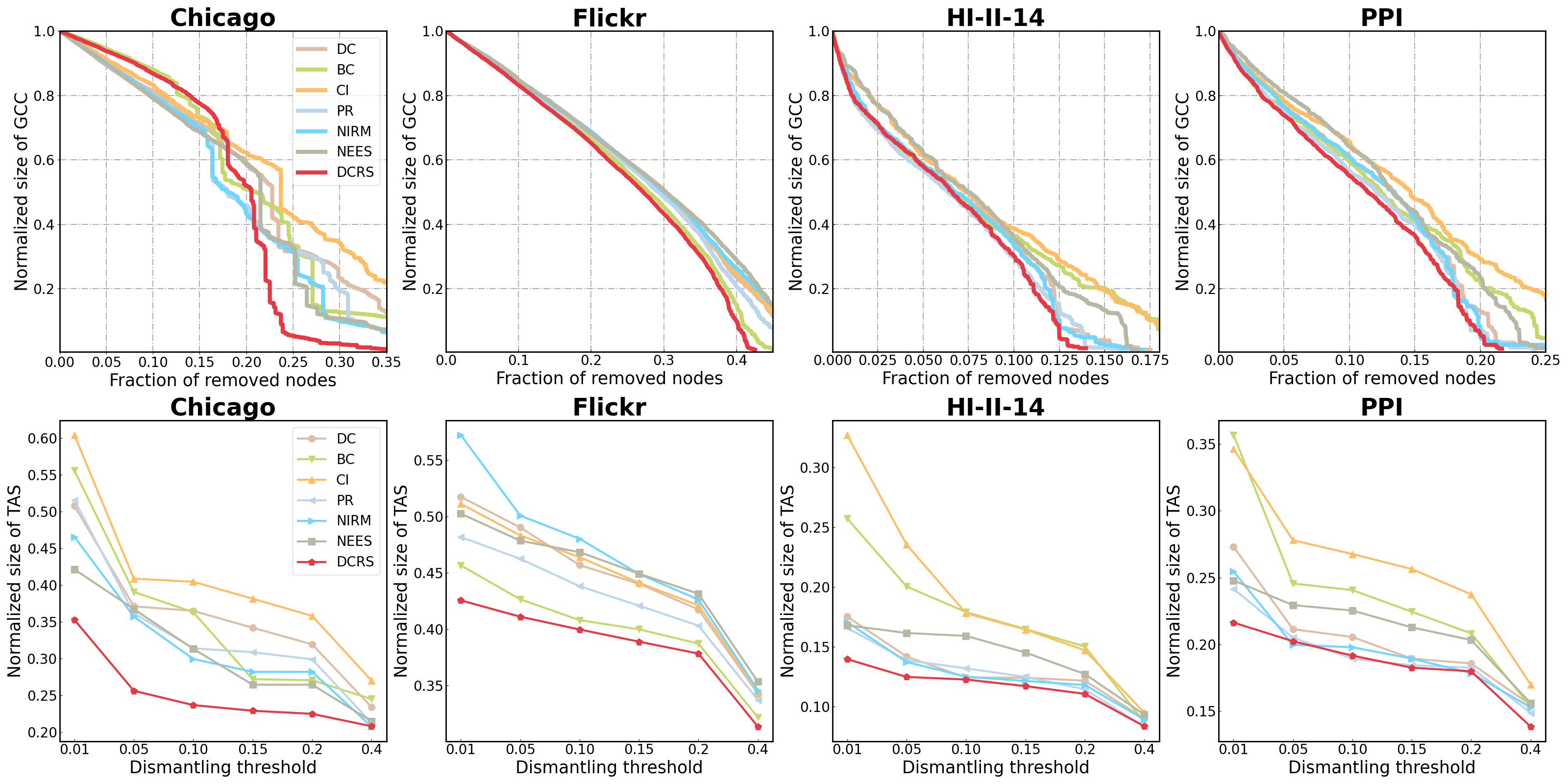}
	\includegraphics[width=0.85\textwidth, height= 0.36\textwidth]{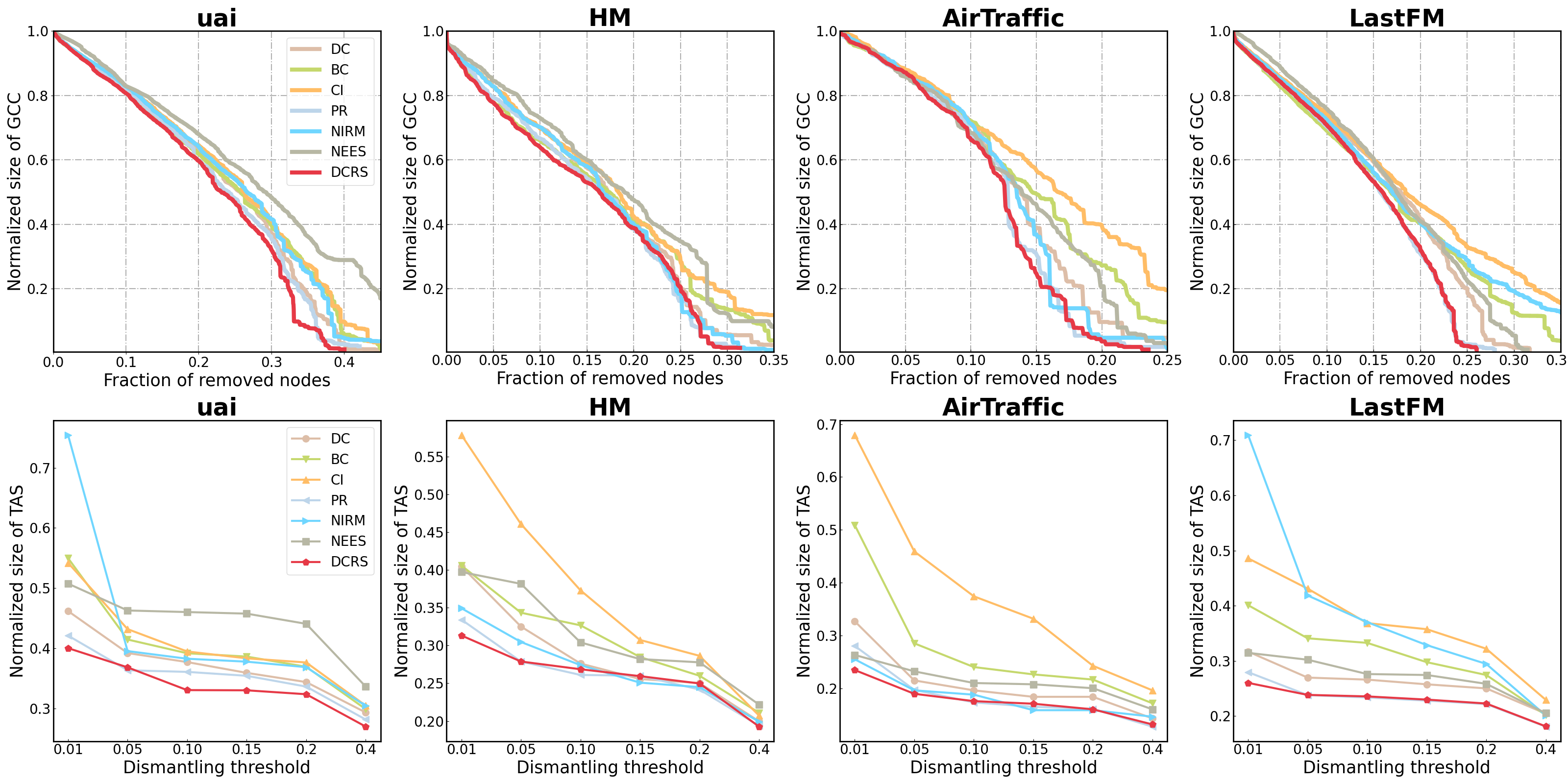}
	\caption{Comparison of dismantling performance on different real-world networks.}
	\label{fig:SI_Real_one}
\end{figure*}

\begin{figure*}[htb]
	\centering
	\includegraphics[width=0.9\textwidth, height= 0.4\textwidth]{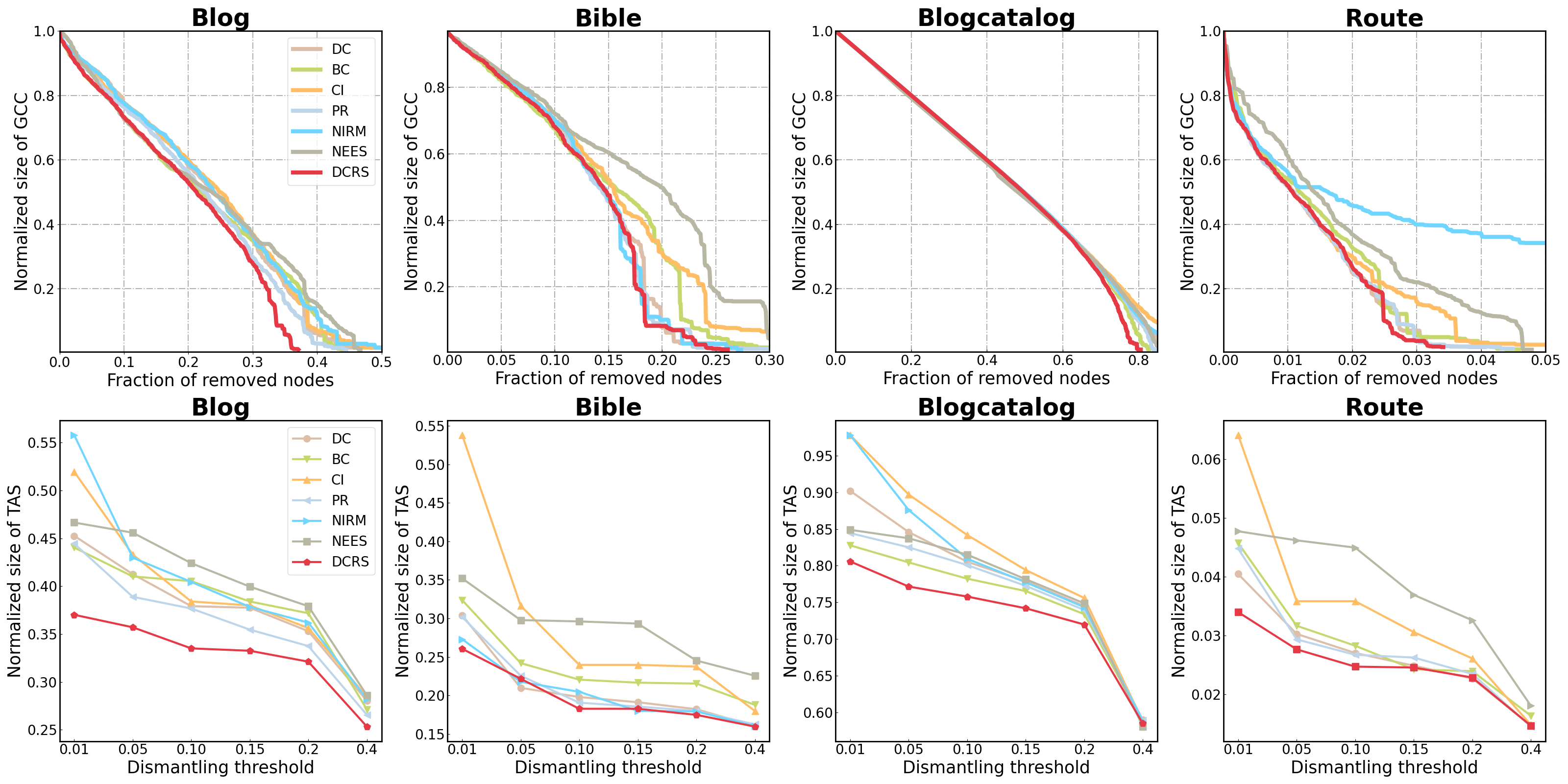}
	\includegraphics[width=0.9\textwidth, height= 0.4\textwidth]{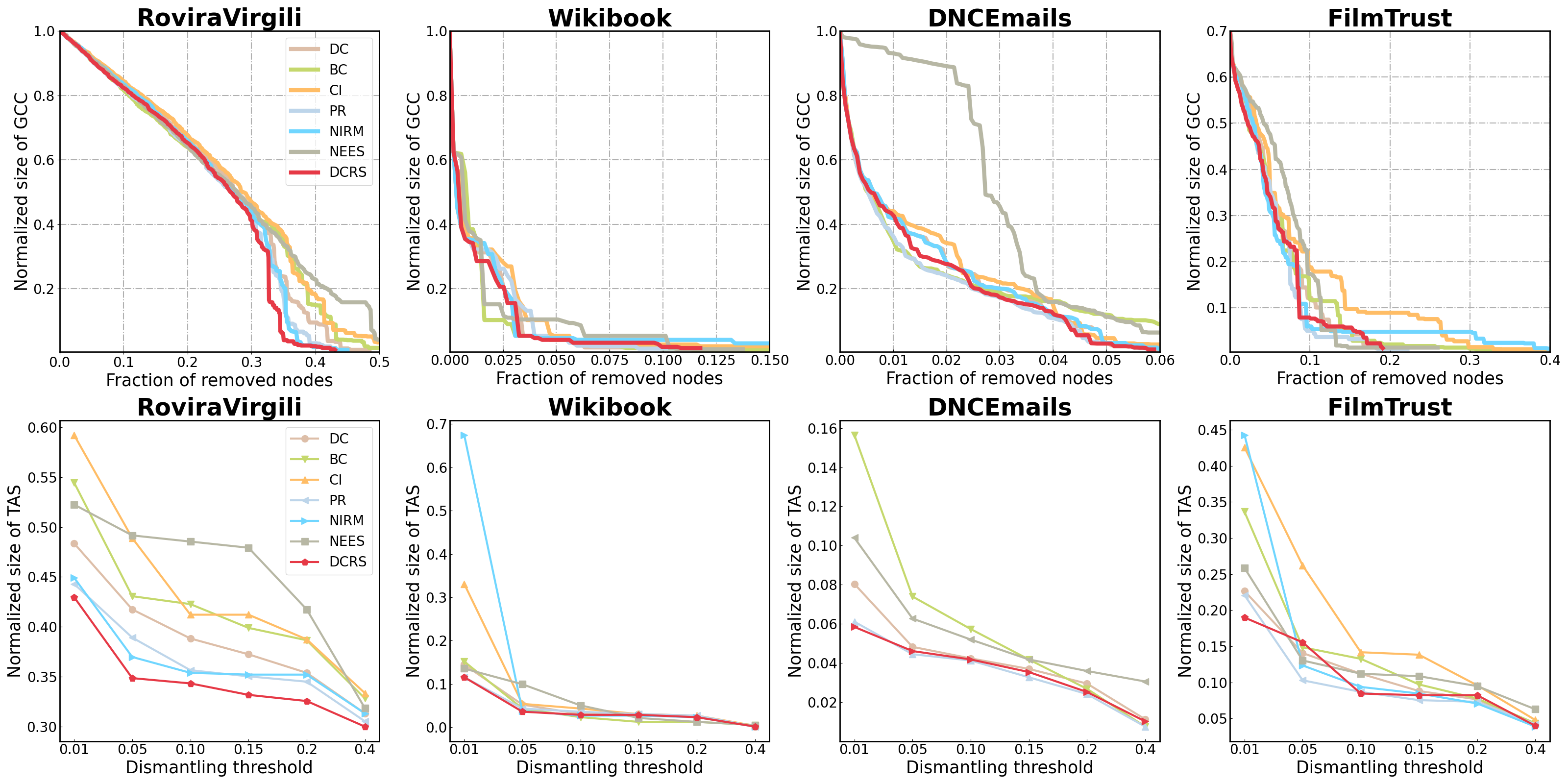}
	\includegraphics[width=0.9\textwidth, height= 0.4\textwidth]{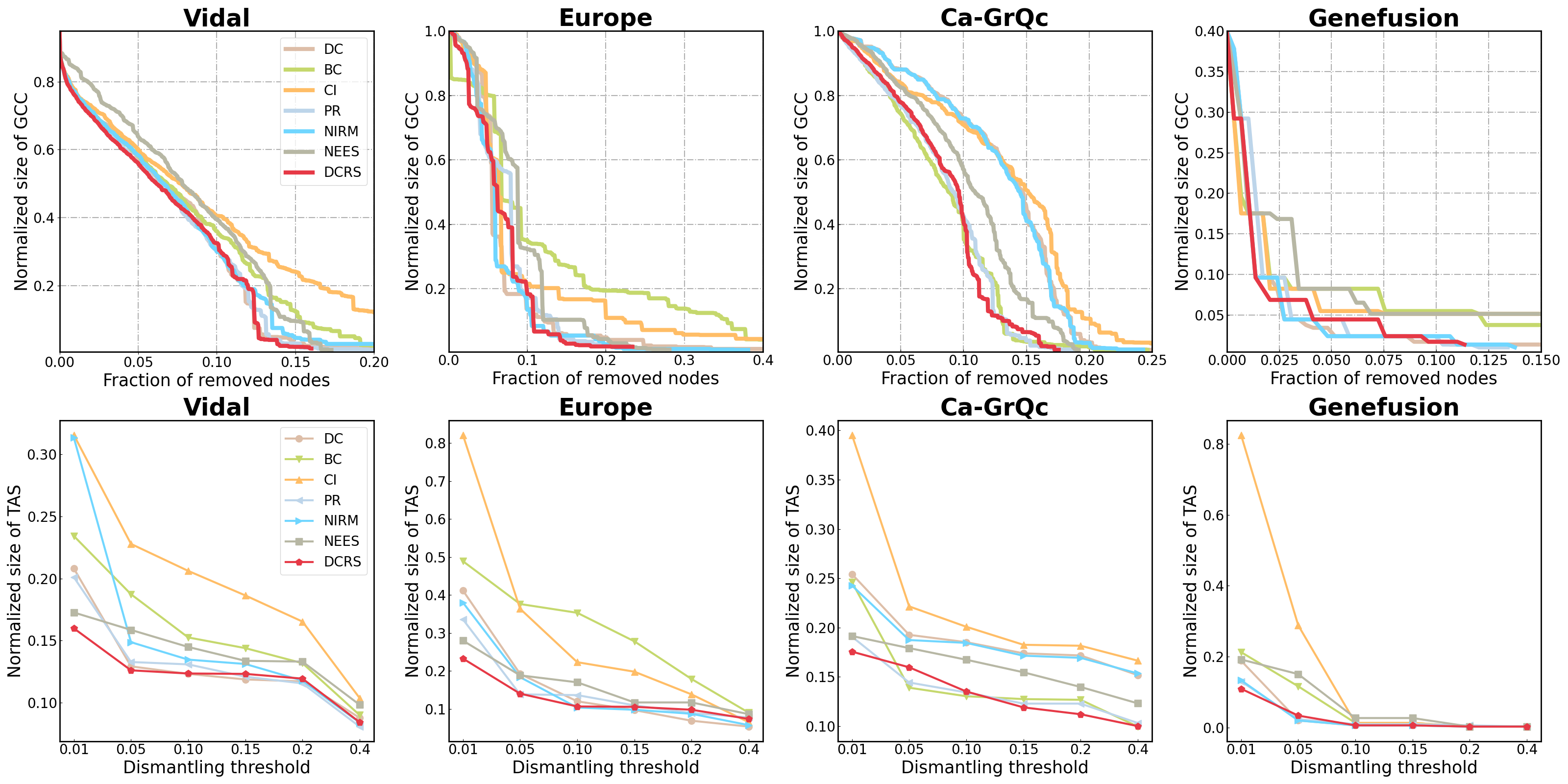}
	\caption{Comparison of dismantling performance on different real-world networks.}
	\label{fig:SI_Real_two}
\end{figure*}

\section*{Appendix}

\subsection*{Results on Synthetic Networks}
For Synthetic networks, we not only compare the results on four classical synthetic models, including BA ($m=4$), ER ($p=6$), WS ($m=8, p=0.8$), PLC ($m=3, p=0.1$) for size $N=1000$ in Fig.~\ref{fig:Syn_Experiment}, but also on different synthetic generation parameters. Specifically, the experiment results on BA networks (also including WS and PLC) with different network sizes $N$ and $m$ are presented in Fig.~\ref{fig:SI_Synthetic}. Our DCRS outperforms the state-of-the-art competitors in terms of smaller TAS size in most cases. Note that we report average results and standard deviations over 20 instances for each synthetic model of the same parameters.

\subsection*{Results on Real-world Networks}
We compare the dismantling performance on twenty two real-world networks from various domains:

\begin{itemize}[leftmargin=*]
	\item \textbf{Chicago~\cite{Eash:Chon:1979:Transportation}}, road transportation network of Chicago region.
	
	\item \textbf{Europe~\cite{Vsubelj:et.al:2011:EurPyhJourB}}, road network located mostly in Europe.
	
	\item \textbf{AirTraffic~\cite{Kunegis:et.al:2013:www}}, constructed from the USA's Federal Aviation Administration National Flight Data Center.
	
	\item \textbf{Gnutella~\cite{Ripeanu:et.al:2002:IEEEComputing}}, which collects Gnutella hosts from 2002.
	
	\item \textbf{Route}~\cite{Leskovec:et.al:2007:TKDD} , the autonomous system networks of the Internet.
	\item \textbf{Blog}~\cite{Kunegis:et.al:2013:www}, which contains front-pages between blogs in the context of 2004 US election.
	
	\item \textbf{FilmTrust}~\cite{Guo:Yorke-Smith:2013:IJCAI}, user–user trust network of the FilmTrust project.
	
	\item \textbf{LastFM~\cite{Rozemberczki:Sarkar:2020:CIKM}}, a social network on Asian LastFM.
	
	\item \textbf{Flickr~\cite{Meng:et.al:2019:WSDM}}, social network through daily sharing and messages on the Flickr website.
	
	\item \textbf{BlogCatalog~\cite{Meng:et.al:2019:WSDM}}, social network with bloggers and relationships from the BlogCatalog website.
	
	\item \textbf{DNCEmails~\cite{Kunegis:et.al:2013:www}}, which is the network of emails in the 2016 Democratic National Committee email leak.
	
	\item \textbf{HM}~\cite{Kunegis:et.al:2013:www}, which contains friendships between users of the website hamsterster.com.
	
	\item \textbf{RoviraVirgili}~\cite{Guimera:et.al:2003:PhyRevE}, email communication network at the University Rovira i Virgili in Tarragona.
	
	\item \textbf{HI-II-14~\cite{Thomas:et.al:2014:Cell}}, a protein interaction network.
	
	\item \textbf{Vidal~\cite{Rual:et.al:2005:Nature}}, a proteome-scale network of proteins.
	
	\item \textbf{Figeys~\cite{Ewing:et.al:2007:MSB}}, a spectral protein interaction network.
	
	\item \textbf{PPI~\cite{Kunegis:et.al:2013:www}}, an organic protein network.
	
	\item \textbf{Genefusion~\cite{Hoglund:et.al:2006:Oncogene}}, which contains interactions between genes during the emergence of cancer.

	\item \textbf{Bible~\cite{Kunegis:et.al:2013:www}}, which contains nouns (places and names) of the King James Version of the Bible and their co-occurrences.
	
	\item \textbf{Wikibook}~\cite{Kunegis:et.al:2013:www}, an edit network of the Kurdish Wikibooks.

	\item \textbf{Ca-GrQc~\cite{Leskovec:Krevl:2014:SNAP} }, which contains papers on Arxiv GR-QC (General Relativity and Quantum Cosmology) from 1993 to 2003.
	
	\item \textbf{UAI~\cite{Wang:et.al:2018:PAKDD}}, a classical network dataset for community detection.
	
\end{itemize}

\par
Table~\ref{tab:SI_Efficiency} compares the attack efficiency (area under the NGCC curve), DCRS achieves the best in 17 out of 22 real networks, and the second best in 5 networks.
Figs.~\ref{fig:SI_Real_one} and \ref{fig:SI_Real_two} present the experiment results on these twenty-two real-world networks, including the normalized size of GCC when removing different fractions of nodes, and the normalized size of TAS against different dismantling thresholds $\Theta$. It can be observed that the proposed DCRS achieves the superior attack efficiency on most of experimented networks.

\end{document}